\documentclass[12pt,english]{article}
\usepackage[T1]{fontenc}
\usepackage[latin1]{inputenc}
\usepackage{a4wide}
\usepackage{graphicx}

\makeatletter

\newcommand{\noun}[1]{\textsc{#1}}

 \newcommand{\lyxaddress}[1]{
   \par {\raggedright #1 
   \vspace{1.4em}
   \noindent\par}
 }

\usepackage{babel}
\makeatother
\begin{document}

\title{Stochastic model of transcription factor-regulated gene expression}

\author{Rajesh Karmakar and Indrani Bose}

\maketitle

\lyxaddress{Department of Physics, Bose Institute, 93/1, A. P. C. Road, Kolkata-700009,
India}

\begin{abstract}
We consider a stochastic model of transcription factor (TF)-regulated
gene expression. The model describes two genes: Gene A and Gene B
which synthesize the TFs and the target gene proteins respectively.
We show through analytic calculations that the TF fluctuations have
a significant effect on the distribution of the target gene protein
levels when the mean TF level falls in the highest sensitive region
of the dose-response curve. We further study the effect of reducing
the copy number of Gene A from two to one. The enhanced TF fluctuations
yield results different from those in the deterministic case. The
probability that the target gene protein level exceeds a threshold
value is calculated with a knowledge of the probability density functions
associated with the TF and target gene protein levels. Numerical simulation
results for a more detailed stochastic model are shown to be in agreement
with those obtained through analytic calculations. The relevance of
these results in the context of the genetic disorder haploinsufficiency
is pointed out. Some experimental observations on the haploinsufficiency
of the tumour suppressor gene, $Nkx\,3.1$, are explained with the
help of the stochastic model of TF-regulated gene expression.
\end{abstract}

\section{Introduction}

Transcription factors (TFs) are proteins which are involved in regulating
gene expression in eukaryotes. The genetic code provides the blueprint
for gene expression, i.e., protein synthesis. Proteins and their complexes
perform several essential functions in the living organism. The TFs
bind at the appropriate regions of the target gene and regulate its
expression by activating/inhibiting the first step in gene expression,
namely, transcription \cite{key-1}. Gene expression involves a series
of biochemical reactions/events which are inherently probabilistic
in nature. Several recent studies, both theoretical and experimental,
highlight the significant influence of stochasticity on gene expression
and its regulation \cite{key-2}. Stochasticity gives rise to fluctuations
around the mean protein level. This is also true for the TFs which
are synthesized by specific genes. In this context, an issue of particular
interest is how the TF fluctuations affect the expression of the regulated
(target) gene. The TFs constitute the {}``dose'' or {}``input signal''
which induces a nonlinear response measured in terms of the amount
of proteins synthesized by the target gene. The dose-response curve,
depicting steady state average values, is in general a sigmoid with
maximum steepness at intermediate levels (region of highest signal
sensitivity) of the input signal. Experiments on synthetic transcription
cascades in \emph{S. cerevisiae} and \emph{E. coli} show that the
effect of TF fluctuations on the target gene protein levels is the
most prominent when the mean TF level falls in the region of highest
signal sensitivity \cite{key-3,key-4}. 

One aspect of the TF-regulated gene expression which is not well explored
relates to the consequences of reducing the copy number of the regulating,
i.e., the TF-synthesizing gene. Eukaryotes, which include higher organisms,
are characterized as diploids in which the set of genes in a cell
has two copies. Haploids, in contrast, have single gene copies. If
one of the copies of a specific gene in an eukaryotic cell is mutated,
i.e., the gene copy number is reduced, the amount of proteins synthesized
is diminished. The functional activity of proteins is often linked
to the requirement that the protein amount cross a threshold level.
If the diminished protein level falls below the threshold, the protein
function is hampered. The loss of a vital protein function may in
certain instances give rise to haploinsufficiency which includes several
genetic diseases \cite{key-5,key-6,key-7}. In TF haploinsufficiency,
one copy of the gene synthesizing the TFs is mutated and the lower
amount of TFs is not sufficient for the functional activity of the
proteins synthesized by the target gene. The p53 tumour suppressor
gene is a case in point. In normal cells, the level of p53 proteins
is low. On DNA damage or under genotoxic stress, the p53 proteins
are activated. These proteins function as TFs and initiate the expression
of several target genes resulting in the activation of a number of
pathways. There are two possible outcomes: either the DNA damage is
repaired (the cell cycle progression is halted temporarily for this
purpose) or if that is not possible, apoptosis, i.e., programmed cell
death occurs. In the absence of any of these outcomes, there is a
proliferation of cells containing damaged DNA through repeated rounds
of the cell division cycle. This triggers the formation and growth
of tumours which may ultimately lead to cancer. Recent experiments
\cite{key-8} show that the mutation of only one of the copies of
the p53 tumour suppressor gene is in many cases sufficient to give
rise to cancer. The cancer is thus a result of TF haploinsufficiency.
Ghosh and Bose \cite{key-9} have shown through model calculations
that in response to DNA damage the transition from the G2 to the mitotic
phase of the cell cycle is delayed, i.e., the cell cycle is arrested
temporarily. This happens when the copy number of the $p53$ tumour
suppressor gene is two. The cell cycle is, however, not arrested when
the copy number is reduced to one. In this case, the DNA damage is
not repaired and a proliferation of cells containing damaged DNA occurs.
In the last few years, several examples of TF haploinsufficiency,
involving a number of tumour suppressor genes, have appeared in the
literature \cite{key-9,key-10,key-11,key-12,key-13}. The proteins
synthesized by the tumour suppressor genes function as TFs and regulate
the expression of the target genes which further activate the relevant
signalling pathways. As in the case of the p53 gene, a reduced gene
dosage leads in certain instances to a loss in the desired outcome,
i.e., the arrest of cell proliferation or apoptosis, resulting in
the formation and growth of tumours. In the case of prostate cancer,
tumour initiation is often brought about by the mutation of one copy
of the tumour suppressor gene $Nkx\,3.1$. There is now considerable
experimental evidence to support the view that both reduced gene copy
number (dosage) and stochasticity in gene expression are essential
factors for the inactivation of the $Nkx\,3.1$- regulated pathways
in a fraction of affected cells \cite{key-11}. 

\begin{figure}
\begin{center}\includegraphics[%
  width=2.5in]{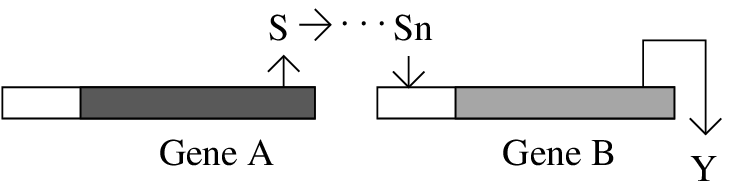}\end{center}

Figure 1. TF-regulated gene expression. Gene A synthesizes the TFs,
$S$ proteins, which regulate the synthesis of $Y$ proteins from
Gene B. $S_{n}$ is the bound complex of $n$ TFs.
\end{figure}

Cook et al. {[}5{]} have considered a minimal model of gene expression
and shown that stochasticity may be an important contributing factor
in the occurrence of haploinsufficiency. They have obtained a number
of interesting results on the stochastic origins of haploinsufficiency
via numerical simulation. When the gene copy number is reduced, the
fluctuations around the mean protein level are enhanced. This sometimes
results in transient excursions of the protein level below the threshold
for the onset of protein activity. In the model of TF-regulated gene
expression, the only stochasticity that is taken into account is that
associated with the expression of the regulated gene. The TFs are
assumed to be present in constant amounts. In this paper, we propose
a stochastic model of TF-regulated gene expression in which Gene A
synthesizes TFs (S proteins) which in turn activate the expression
of Gene B leading to the synthesis of Y proteins (figure 1). The concentrations
of the respective proteins are also denoted by S and Y. In our model,
the expressions of both Gene A and Gene B are considered to be stochastic
in nature. Using an analytical formalism, we explore how the fluctuations
in the amount of TFs affect the distribution of the Y protein levels.
We further study the effect of a reduction in the copy number of Gene
A from two to one (the case of TF haploinsufficiency) on the expression
of the target gene (Gene B). For simplicity, the copy number of Gene
B is assumed to be one. The results from the analytical calculation
are shown to be consistent with those obtained through stochastic
simulations based on the Gillespie algorithm \cite{key-27}. Some
experimental observations on the haploinsufficiency of the tumour
suppressor gene, $Nkx\,3.1$, are explained using the stochastic model
of TF-regulated gene expression.

\section{Stochastic Model}

In the minimal model of stochastic gene expression \cite{key-5},
a gene can be in two possible states: inactive $(G)$ and active $(G^{*})$.
Random transitions occur between the states $G$ and $G^{*}$ according
to the Scheme (Scheme 1),\begin{equation}
\begin{array}{ccccccc}
 & k_{a} &  & j_{p} &  & k_{p}\\
G & \rightleftharpoons & G^{*} & \longrightarrow & p & \longrightarrow & \Phi\\
 & k_{d}\end{array}\label{eq:1}\end{equation}

\noindent where $k_{a}$ and $k_{d}$ are the stochastic activation
and deactivation rate constants. In the active state $G^{*}$, the
gene synthesizes a protein $(p)$ with the rate constant $j_{p}$.
The protein product degrades with a rate constant $k_{p}$, the degradation
product being represented by $\Phi$. The model (equation (1)) describes
constitutive gene expression. In the case of eukaryotes, activation
from the state $G$ to the state $G^{*}$ is brought about by the
binding of a complex $S_{n}$ of $n$ individual TFs to the gene.
Gene expression now takes place according to the Scheme (Scheme 2)\begin{equation}
\begin{array}{ccccccccc}
 & k_{1} &  & k_{a}^{'} &  & j_{p}^{'} &  & k_{p}^{'}\\
G+S_{n} & \rightleftharpoons & GS_{n} & \rightleftharpoons & G^{*} & \longrightarrow & p & \longrightarrow & \Phi\\
 & k_{2} &  & k_{d}^{'}\end{array}\label{eq:2}\end{equation}

\noindent where $GS_{n}$ represents the bound complex of $G$ and
$S_{n}$. Though the Schemes 1 and 2 do not fully capture the complexity
of gene expression, they include the essential features. The simple
schemes provide the theoretical framework for gaining important insight
on stochastic gene expression and also for interpreting experimental
results \cite{key-2}. As has been shown earlier \cite{key-5,key-15,key-16},
the Schemes 1 and 2 are equivalent with effective rate constants $k_{a}^{''}$
and $k_{d}^{''}$ given by\begin{equation}
k_{a}^{''}=k_{a}^{'}\frac{(S/K)^{n}}{1+(S/K)^{n}}\;\;,\qquad k_{d}^{''}=k_{d}^{'}\label{eq:3}\end{equation}

\noindent where $K^{n}=\frac{k_{2}}{k_{1}}\: K_{n}$ , $K_{n}$ being
the equilibrium dissociation constant for the reaction $n\: S\rightleftharpoons S_{n}$.
The effective activation rate constant $k_{a}^{''}$ has the form
of a Hill function.

We now consider the stochastic model corresponding to the Scheme 1.
In the model, the only stochasticity arises from the random transitions
of a gene between the active and inactive states as in the minimal
model of Cook et al. \cite{key-5}. Protein synthesis and degradation
occur in a deterministic manner. In each state of the gene, the concentration
$X$ of proteins evolves according to the equation\begin{equation}
\frac{dX}{dt}=j_{p}z-k_{p}X\label{eq:4}\end{equation}

\noindent where $z=1(0)$ when the gene is in the active (inactive)
state. 

\noindent Let $p(X)$ be the probability density function (PDF) of
the protein levels in the steady state. This is given by \cite{key-15}

\noindent \begin{equation}
p(X)=C\,(k_{p}X)^{r_{1}-1}(j_{p}-k_{p}X)^{r_{2}-1}\label{eq:5}\end{equation}

\noindent where $r_{1}=k_{a}/k_{p},$ $r_{2}=k_{d}/k_{p}$ and $C$
is the normalization constant.

In the model of TF-regulated gene expression, the synthesis of the
TFs ($S$ proteins) and that of the $Y$ proteins occur according
to the Schemes 1 and 2 respectively. The different rate constants
in the two cases are as specified in equations (1)-(3). In the deterministic
formalism, the steady state concentrations of the TFs and the $Y$
proteins are given by \begin{equation}
S_{mean}=\frac{n_{G}\: j_{p}}{k_{p}}\,\frac{k_{a}}{k_{a}+k_{d}}\;,\;\; Y_{mean}=\frac{j_{p}^{'}}{k_{p}^{'}}\:\frac{k_{a}^{''}}{k_{a}^{''}+k_{d}^{''}}\label{eq:6}\end{equation}

\noindent where $n_{G}$ is the copy number of Gene A. Let $Y_{thr}$
be the threshold level for the onset of activity of the $Y$ proteins.
If $Y_{mean}$ is $<Y_{thr}$, a loss in the protein function occurs.
If $Y_{mean}$ is $>Y_{thr}$, the normal protein activity is not
hampered. With the stochastic expression of both the Genes A and B
taken into account, the steady state distributions of the TF and the
$Y$ protein levels are spread around the mean values $S_{mean}$
and $Y_{mean}$. Let $q(S)$ and $Q(S)$ be the PDFs of the TF levels
when the copy number of Gene A is $1$ and $2$ respectively. The
corresponding PDFs of the $Y$ proteins are $p_{1}(Y)$ and $p_{2}(Y).$
If the distribution of the $Y$ protein levels overlaps with $Y_{thr}$,
the steady state probability, $p(Y<Y_{thr})$ $(p(Y>Y_{thr}))$, that
the protein level $Y$ is $<$ $Y_{thr}$ ( >
 $Y_{thr}$), is non-zero even if $Y_{mean}$ is $>$ $Y_{thr}$ ($<$$Y_{thr}$).

We now study the effect of TF fluctuations on the distribution of
the Y protein levels. We consider one copy each of the Genes A and
B and assume the parameter values to be $k_{a}=k_{d}=4,$ $j_{p}=1000,$
$k_{p}=1$ (Gene A) and $k_{a}^{'}=12,$ $k_{d}^{'}=4,$ $j_{p}^{'}=1000,$
$k_{p}^{'}=1$ (Gene B) in appropriate units. The parameter $K$ is
chosen to be $K=500$ so that $S_{mean}=K$ ($S_{mean}$ is given
by equation (6) with $n_{G}=1$). Let $p_{1}(Y,S)$ be the steady
state PDF of the $Y$ protein levels for a fixed amount $S$ of TFs.
The PDF has the same form as in equation (5) but with $k_{a}$, $k_{d}$,
$j_{p}$ and $k_{p}$ replaced by $k_{a}^{''}$, $k_{d}^{''}$, $j_{p}^{'}$
and $k_{p}^{'}$. Similarly, the PDF $q(S)$ has the form as in equation
(5). The steady state PDF is given by\begin{equation}
p_{1}(Y)=\int_{all\: S}p_{1}(Y,S)\, q(S)\: dS\label{eq:8}\end{equation}

\noindent Since $q(S)$ and $p_{1}(Y,S)$ are known analytically,
$p_{1}(Y)$ can be calculated using Mathematica. Figure 2(a) shows
the variation of $k_{a}^{''}$ (equation (3)) with $S$ for different
values of the Hill coefficient, $n=1$ (curve {}``a''), $n=4$ (curve
{}``b'') and $n=12$ (curve {}``c''). Figure 2(b) shows the steady
state distribution $p_{1}(Y)$ versus $Y$ for $n=1$ (curve {}``a''),
$n=4$ (curve {}``b'') and $n=12$ (curve {}``c''). The curve
{}``d'' describes the distribution of $p_{1}(Y,S_{mean})$ with
$S$ fixed at $S_{mean}$, the mean amount of the TFs in the steady
state i.e., the TF fluctuations are ignored. From figure 2(b), one
finds that as the value of the Hill coefficient $n$ increases, the
effect of the TF fluctuations on the distribution of $p_{1}(Y)$ becomes
more and more prominent. The reason for this is not difficult to find.
For $n>1$, the sharpest change in the $k_{a}^{''}$ versus $S$ curve
(equation (3)) occurs around $S\sim K$. Figure 2 has been obtained
for $K=S_{mean}=500$ so that for large $n$ even small fluctuations
around $S_{mean}$ give rise to considerable changes in the value
of the effective rate constant $k_{a}^{''}$. The region close to
this point defines the highest signal sensitive region of the dose-response
curve. 

\begin{figure}
\includegraphics[%
  width=2.5in]{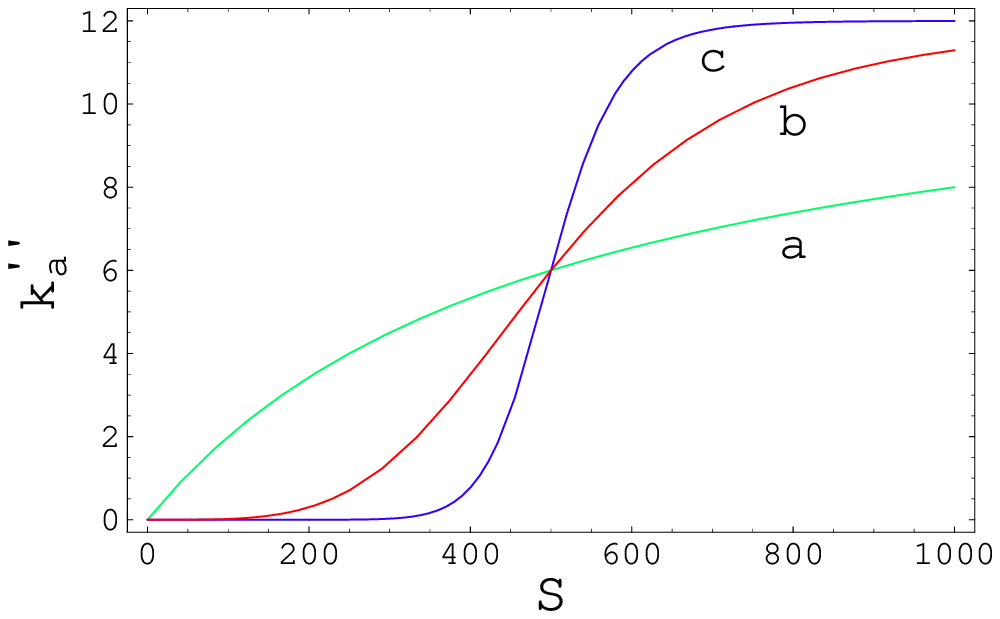} \includegraphics[%
  width=2.5in]{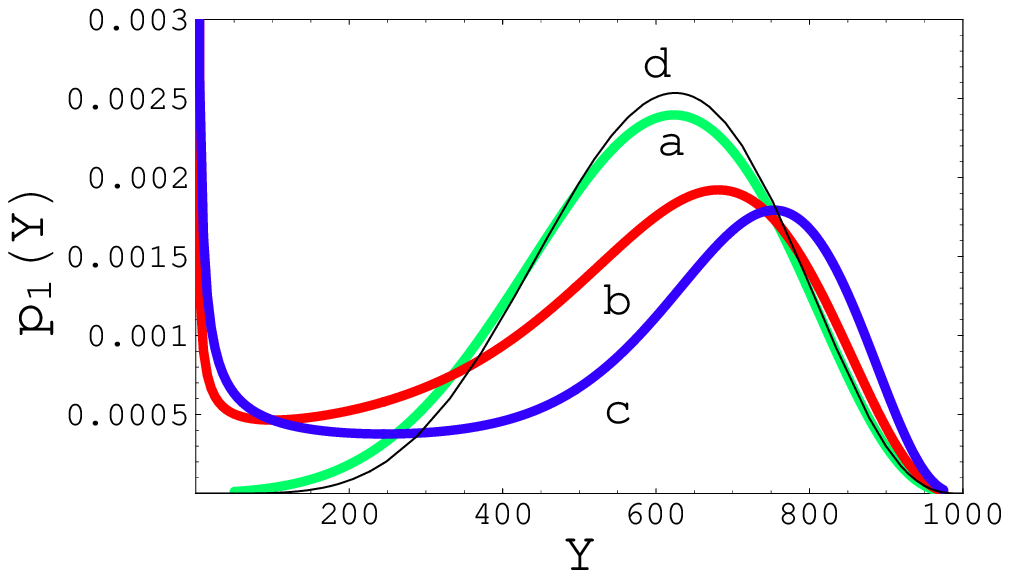}

\begin{center}(a)$\qquad\qquad\qquad\qquad\qquad\qquad\qquad$(b)\end{center}

Figure 2. (a) Variation of the effective rate constant $k_{a}^{''}$as
a function of $S$ (equation (3)) for different Hill coefficients,
$n=1$ (curve {}``a''), $n=4$ (curve {}``b''), $n=12$ (curve
{}``c''), (b) The distribution $p_{1}(Y)$ versus $Y$ of protein
levels for different Hill coefficients, $n=1$ (curve {}``a''),
$n=4$ (curve {}``b'') and $n=12$ (curve {}``c''). Curve {}``d''
curve describes the distribution for a fixed amount, $S_{mean}$,
of TFs.
\end{figure}

\begin{figure}
\includegraphics[%
  width=2.5in]{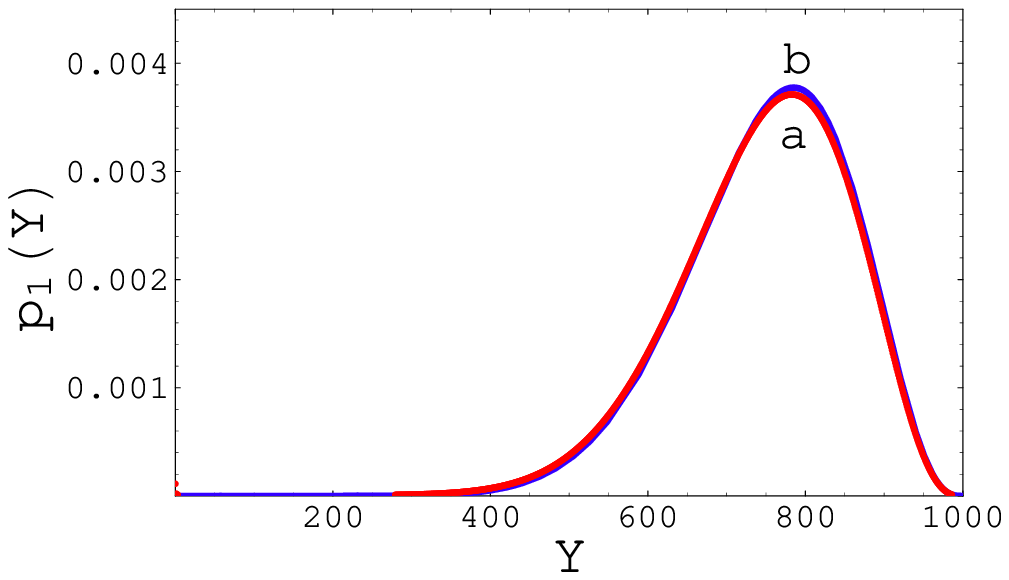} \includegraphics[%
  width=2.5in]{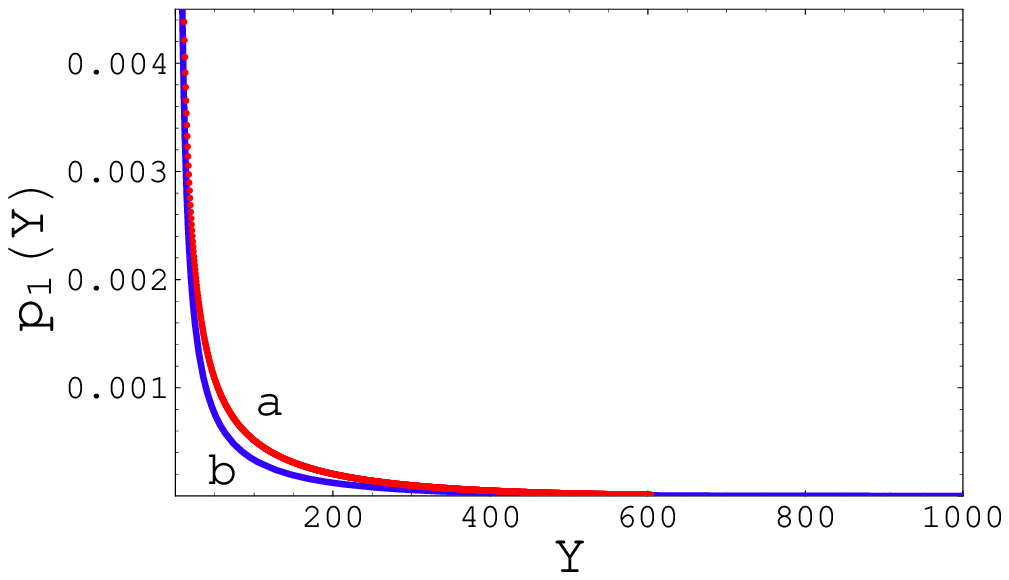}

\begin{center}(a)$\qquad\qquad\qquad\qquad\qquad\qquad\qquad$(b)\end{center}

Figure 3. Steady state distribution of the $Y$ protein level for
(a) $S_{mean}/K=4$ and (b) $S_{mean}/K=0.25$. Curve {}``a'' ({}``b'')
is obtained with the TF fluctuations taken (not taken) into account.
\end{figure}

For $S_{mean}/K<<1$ or $>>1$, the TF fluctuations have a less marked
effect as the change in the value of $k_{a}^{''}$ is not as much
as in the case when $S_{mean}/K\sim1$. This is shown in figures 3(a)
and (b). The parameter values are the same as in the case of figure
2 with the Hill coefficient $n=4$ except that $S_{mean}/K=4$ (figure
3(a)) and $S_{mean}/K=0.25$ (figure 3(b)). Curve {}``a'' ({}``b'')
describes the steady state distribution $p_{1}(Y)$ versus $Y$ with
the TF fluctuations taken (not taken) into account. The findings from
figures 2 and 3 are consistent with the earlier observation \cite{key-2,key-3}
that the effect of TF fluctuations on the target gene protein levels
is most clearly seen in the region of highest signal sensitivity ($S_{mean}/K\sim1$).
Figure 2(b) ($n=4$ and $n=12$) shows that the TF fluctuations give
rise to a bimodal distribution in the $Y$ protein levels. This is
in agreement with the experimental observations of Blake et al. \cite{key-3}.
We also find that the higher the value of $n$, the greater is the
range of values of $S_{mean}/K$ for which the TF fluctuations can
be ignored. Furthermore, with lesser fluctuations in the TF levels
(narrower distribution), the effect of the TF fluctuations on the
$Y$ protein levels becomes significant for higher values of $n$.
This is demonstrated in figure 4. A narrower distribution in the TF
protein levels, than in the cases of figures 2 and 3, is obtained
with the choice of the parameter values $k_{a}=k_{d}=40,$ $j_{p}=1000,$
$k_{p}=1$. The target gene has the parameter values $k_{a}^{'}=12,$
$k_{d}^{'}=4,$ $j_{p}^{'}=1000,$ $k_{p}^{'}=1$ and $S_{mean}=K=500$.
Figure 4(a) shows that for the Hill coefficient $n=4$, the $p_{1}(Y)$
versus $Y$ curve is little affected by the TF fluctuations even when
$S_{mean}/K\sim1$. The effect becomes prominent for higher values
of $n$ (figure 4(b), $n=12$). 

The major result obtained in this section is to demonstrate through
analytical calculations that the TF fluctuations may considerably
affect the distribution of the target gene protein levels. This is
particularly so when the mean TF level falls in the highest sensitive
region of the dose-response ($k_{a}^{''}$ vs $S$ ) curve describing
the activation of the target gene. The result, though based on a specific
set of parameter values, illustrates a general feature of TF-regulated
gene expression. In many such cases, the TFs bind the target gene
at multiple sites. This, combined with cooperative interactions between
the TFs, imparts an ultrasensitive character to the dose-response
curve. The same is true if a bound complex of TF molecules binds the
operator region. If the mean TF level falls in the regions around
the steepest part of the curve, the fluctuations around the mean level
give rise to fluctuations in the effective activation rate constant,
$k_{a}^{''}$, of the target gene. For sufficiently strong TF fluctuations,
the distribution of the target gene protein levels is significantly
altered from the case when the TF fluctuations are ignored. There
have been earlier studies \cite{key-5,key-18,key-19}, based on simple
models of TF-regulated gene expression, which did not take the TF
fluctuations explicitly into account. Simpson et al. \cite{key-33}
have carried out a comprehensive analysis of stochasticity in gene
transcriptional regulation based on the frequency domain Langevin
approach. The transcriptional regulation occurs via protein (inducer)-DNA
interactions at an operator site of the target gene. A significant
achievement of the study is to obtain the frequency distribution of
the target gene expression noise and identify the impact of noise
originating from operator fluctuations. The operator fluctuations
constitute an important component of the stochasticity associated
with TF-regulated gene expression. The noise associated with the expression
of a specific gene has two components: intrinsic and extrinsic of
which the latter has the more dominated contribution to the total
noise \cite{key-2}. Recent reports suggest that this is particularly
so for highly expressed genes \cite{key-34,key-35}. Both numerical
computations and experiments \cite{key-2,key-3} show that a major
component of the extrinsic noise stems from the fluctuations in the
TF activity. Our simple stochastic model of TF-regulated gene expression
yields analytical expressions for the PDFs describing the distributions
of the TF and target gene protein levels. This makes it particularly
convenient to study the effect of the TF fluctuations on the expression
of the target gene. 

\begin{figure}
\includegraphics[%
  width=2.5in]{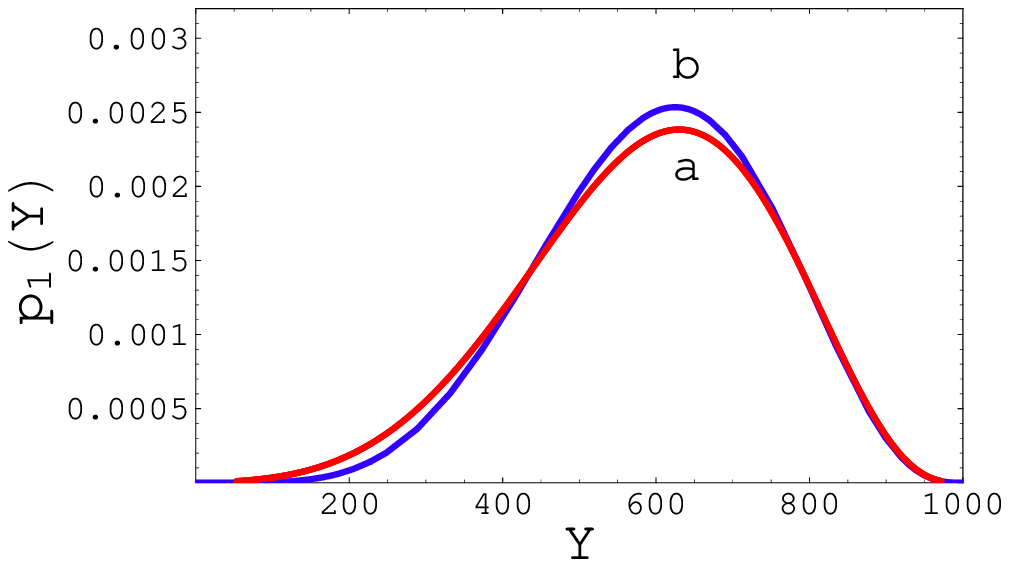} \includegraphics[%
  width=2.5in]{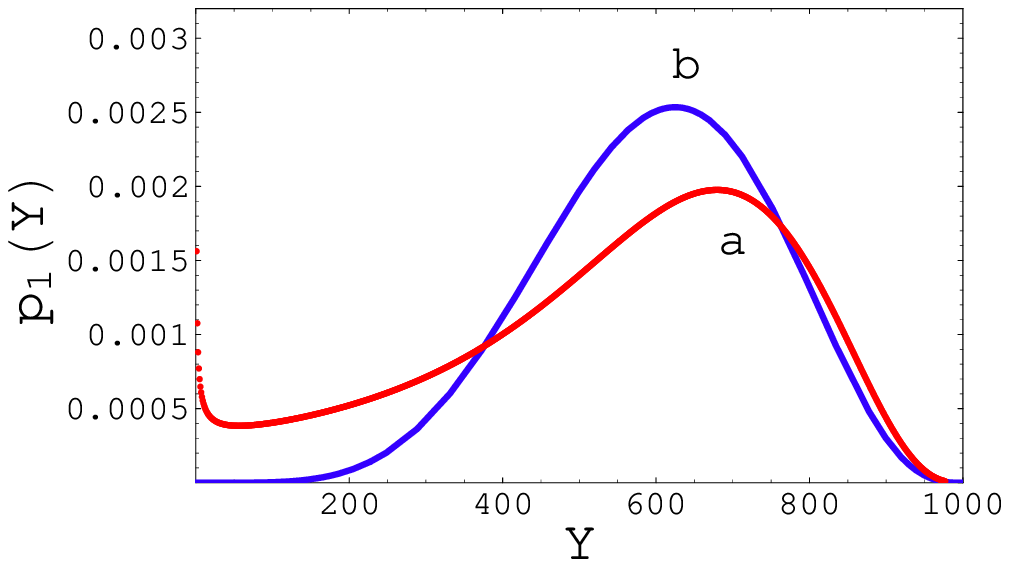}

\begin{center}(a)$\qquad\qquad\qquad\qquad\qquad\qquad\qquad$(b)\end{center}

Figure 4. Steady state distribution of the $Y$ protein levels for
Hill coefficient $n=4$ (a) and $n=12$ (b) respectively. Curve {}``a''
({}``b'') is obtained with the TF fluctuations taken (not taken)
into account. The distribution of the TF levels is less noisy than
that in the case of figure 2.
\end{figure}

The analytical tractability of the model arises from two assumptions.
Firstly, the two major steps of gene expression, namely, transcription
(synthesis of mRNAs) and translation (synthesis of proteins) have
been combined into a single step leading to protein production. Secondly,
the only source of stochasticity in the model lies in the random activation
and deactivation of the target gene expression. The first assumption
provides the basis for several studies of stochastic gene expression
\cite{key-5,key-18,key-19,key-20}. The second assumption is strictly
valid when the dominant source of noise is associated with the random
activation and deactivation of gene expression. This holds true in
the case of large steady state gene product level studied by Kepler
and Elston \cite{key-18}. In this limit, slow promoter kinetics (rates
of gene activation and deactivation lower than the synthesis rate
of the gene product) constitute a major source of noise. Fast transitions
between the promoter states, on the other hand, generate low amounts
of noise. The effect of the TF fluctuations on the target gene expression
is prominent in the case of slow promoter kinetics. As discussed in
detail in Ref. \cite{key-2}, slow promoter kinetics give rise to
increased heterogeneity within a cell population including bimodal
population distributions. Slow transitions between the promoter states
are particularly relevant in the case of eukaryotic gene expression
resulting in transcriptional bursts of mRNA synthesis (the production
of mRNA occurs in pulses) \cite{key-2,key-21}.

\section{Reduced Gene Copy Number}

We next consider the effect of reducing the copy number of Gene A
from two to one on the distribution of target gene protein levels.
Figure 5 shows that a reduction in the copy number $n_{G}$ of Gene
A from two (red curve) to one (blue curve) gives rise to a wider distribution
in the TF levels, i.e., the fluctuations around the mean protein level
are larger. The steady state PDF $Q(S)$ ($n_{G}=2$) is given by\begin{equation}
Q(S)=\int_{all\, s_{1}}q(s_{1})\, q(S-s_{1})\: ds_{1}\label{eq:7}\end{equation}

\noindent where $S$ denotes the total TF concentration with $S=s_{1}+s_{2}$,
the sum of the concentrations of the TFs synthesized by the individual
copies of the Gene A. The PDF $q(s_{i})$ $(i=1,\:2)$ has the form
given in equation (5). The parameter values for both the one-gene
and the two-gene cases are the same as in the case of figure 2 except
that $j_{p}=2000$ in the one-gene case. This has been done to keep
$S_{mean}$ fixed and facilitate the comparison of the two distributions.
A measure of noise is given by $\chi=standard$ $deviation$$/mean$.
If $j_{p}=1000$ in both the one- and two-gene cases, $\chi$ has
the values $\chi=0.236$ (two gene copies) and $\chi=0.333$ (one
gene copy). Figure 6 shows the steady state distributions $p_{1}(Y)$
(curve {}``b'') and $p_{2}(Y)$ (curve {}``a'') when the copy
number of Gene A is $1$ and $2$ respectively. The parameter values
are the same as in the case of figure 2 with the Hill coefficient
$n=4.$ One finds that the distribution of the $Y$ protein levels
is considerably changed when the copy number of the regulating gene
is reduced. In the one-gene case, the value of $S_{mean}$ falls to
$500$ i.e., $S_{mean}/K=1$. In the parameter region close to this
point, the effect of fluctuations is the most prominent. Again, one
finds that the TF fluctuations give rise to a bimodal distribution
in the $Y$ protein levels. 

\begin{figure}
\begin{center}\includegraphics[%
  width=3in]{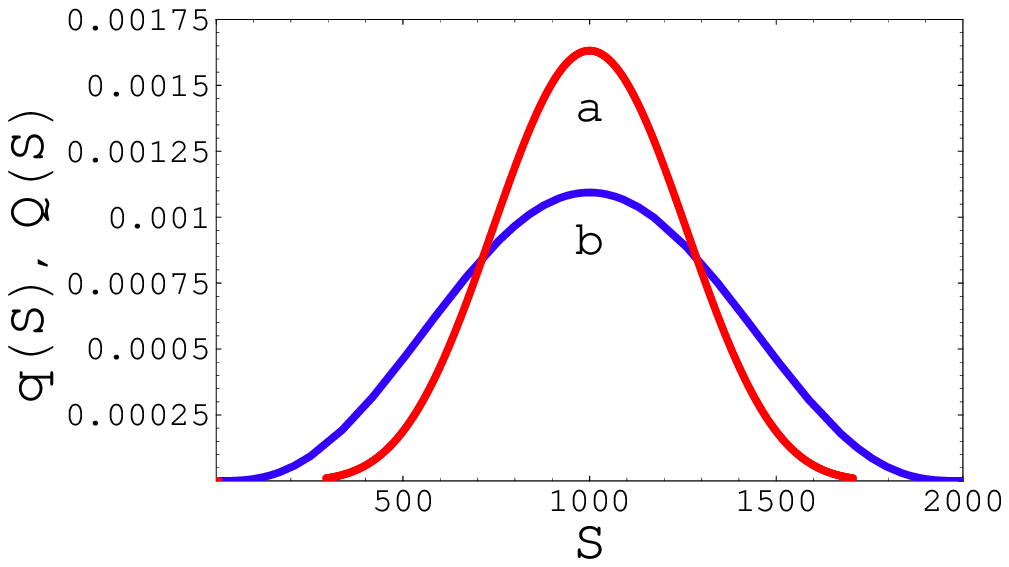}\end{center}

Figure 5. Steady state distribution of protein $Y$ levels when the
copy number of Gene A is two (curve {}``a'') and one (curve {}``b'')
\end{figure}

\begin{figure}
\begin{center}\includegraphics[%
  width=3in]{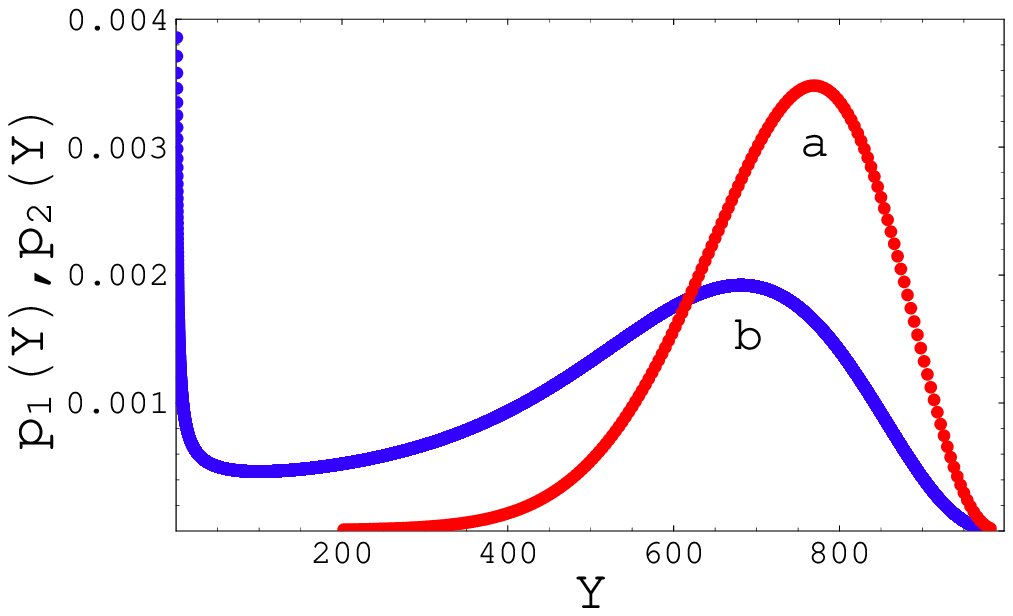}\end{center}

Figure 6. Steady state distribution of protein $Y$ levels when the
copy number of Gene A is two (curve {}``a'') and one (curve {}``b'')
\end{figure}

Figure 7 shows the steady state probability $p(Y>Y_{thr})$ that the
protein level $Y$ exceeds a threshold value, $Y_{thr}$, versus the
activation rate constant $k_{a}$ of Gene A. The dotted (dot-dashed)
curve corresponds to the case when the copy number of Gene A is 2
(1). In the one-gene case,\begin{equation}
p(Y>Y_{thr})=1-\int_{0}^{Y_{thr}}\int_{all\, S}p_{1}(Y,S)\: q(S)\: dS\, dY\label{eq:9}\end{equation}

\noindent The expression for the two-gene case is obtained by substituting
$p_{1}(Y,S)$ and $q(S)$ by $p_{2}(Y,S)$ and $Q(S)$ respectively.
The threshold value $Y_{thr}$ is set at $25$ percent of the maximum
amount $Y_{max}$ of the $Y$ protein when the copy number of Gene
A is two ($Y_{max}=j_{p}^{'}/k_{p}^{'}=1000$ so that $Y_{thr}=250$).
From figure 7 one finds that the reduction of the copy number of Gene
A diminishes the probability of maintaining the output protein level
above a threshold value. Variation of the activation rate constant
$k_{a}$ changes $S_{mean}$, the mean amount of TFs (Eq. (6)) regulating
the expression of Gene B. Figure 7 also includes the curves for $p(Y>Y_{thr})$
when the fluctuations in the TF amounts are ignored, i.e., the TF
amount is kept fixed at $S_{mean}$. The dashed (solid) curve describes
the two-gene (one-gene) case. For the same gene copy number, the curves,
with and without the TF fluctuations, intersect at a value of $k_{a}=k_{ac}$
for which $Y_{mean}=Y_{thr}$. When $k_{a}$ is $<$ $k_{ac}$, $Y_{mean}$
is $<Y_{thr}$. The distribution of the $Y$ protein levels around
$Y_{mean}$ becomes broader when the TF fluctuations are taken into
account, i.e., the TF levels are assumed to have a distribution around
$S_{mean}$. Thus, $p(Y>Y_{thr})$ in this case is larger than when
the TF concentration is kept fixed at $S_{mean}$. When $k_{a}$ is
$>$ $k_{ac}$, $Y_{mean}$ is $>$ $Y_{thr}$. In this case, $p(Y>Y_{thr})$,
with the TF fluctuations taken into account, is smaller than $p(Y>Y_{thr})$,
with the TF concentration kept fixed. One thus has the interesting
result that the TF fluctuations can both augment and diminish the
probability $p(Y>Y_{thr})$. The $p(Y>Y_{thr})$ versus $k_{a}$ curve
is also found to be less steep in the presence of the TF fluctuations.
One can further show that the curve is steeper for higher values of
the Hill coefficient $n.$ 

\begin{figure}
\begin{center}\includegraphics[%
  width=3in]{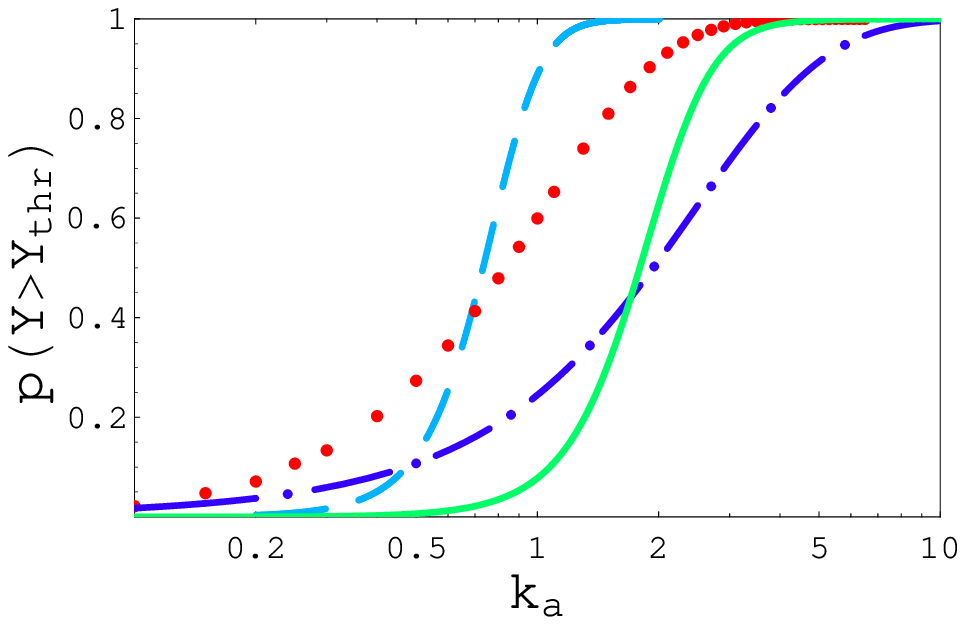}\end{center}

Figure 7. The probability $p(Y>Y_{thr})$ versus the activation rate
constant, $k_{a}$, of Gene A when the copy number of Gene A is two
(dotted and dashed curves) and one (dot-dashed and solid curves).
The dashed and solid curves represent the cases in which the TF fluctuations
are ignored.
\end{figure}

In section 2, we have outlined some arguments for the validity of
our stochastic model. We now show that similar results regarding the
effect of TF fluctuations are obtained when a more detailed stochastic
model is considered. In the model, the expression of both the Genes
A and B take place through two steps: transcription and translation,
i.e., the syntheses of mRNAs and proteins are treated separately.
The production and the degradation of the mRNAs and the proteins are
considered to be stochastic processes. The formation of a complex
of $n$ TF molecules which regulates the expression of the target
gene is also taken to be a stochastic event. With full stochasticity
taken into account, the model is no longer analytically tractable.
We undertake numerical simulation based on the Gillespie algorithm
\cite{key-27} to obtain the distribution of protein $Y$ levels.
The results are displayed in figure 8. The target gene protein levels
have a unimodal distribution when the copy number of the regulating
gene (Gene A) is two (figure 8(a)), a bimodal distribution (figure
8(b)) is obtained when the copy number of Gene A is $1$ (increased
TF fluctuations). The results have been obtained for slow promoter
kinetics leading to the production of mRNAs in bursts. The amount
of proteins synthesized is also not small. A heterogeneous distribution
of the levels of expression, including bimodality, is obtained if
the mRNA and protein half lives are shorter than the average time
between the successive bursts of transcription \cite{key-2,key-15,key-19,key-21}.
The simulation and analytical model results are in good agreement
for low transcriptional burst rates. Higher transcriptional burst
rates demand faster transitions between the inactive and active states
of the gene. In this case, the noise contributed by the promoter activation-inactivation
kinetics is lower. Depending on the gene product level, the noise
associated with transcription and translation may become more dominant
so that the analytical model ceases to be valid.

We now discuss the occurrence of haploinsufficiency, due to the loss
of one copy of the tumour suppressor gene $Nkx\,3.1$. In this case,
haploinsufficiency is manifest in tumour initiation in the prostate
leading to cancer \cite{key-11,key-13}. The tumour suppressor genes
are generally activated under DNA damage or genotoxic stress. The
function of the proteins synthesized by these genes is to limit cell
growth or survival. Mice lacking one copy of such a gene (examples
include the $p27$, $p53$, $Dmp1$ and $Nkx\,3.1$ genes) are known
to develop cancerous or pre-cancerous lesions despite protein synthesis
from the remaining copy of the gene. This indicates a failure in checking
cell proliferation or bringing about cell death. The $Nkx\,3.1$ gene
has several positively and negatively regulated target genes which
exhibit a variety of responses to the loss of one copy of the $Nkx\,3.1$
gene. We consider the examples of two of the positively regulated
genes, probasin and intelectin. Probasin is relatively insensitive
to the loss of one $Nkx\,3.1$ copy, i.e., both the wild-type and
$Nkx\,3.1^{+/-}$prostates witness high levels of probasin expression.
The expression is retained even in $Nkx\,3.1^{-/-}$ (loss of both
gene copies) indicating a basal level of probasin expression. Intelectin
is, however, highly dosage sensitive and is not expressed in either
the $Nkx\,3.1^{+/-}$ or $Nkx\,3.1^{-/-}$ prostate. In situ hybridization
experiments reveal heterogeneous expression patterns for both probasin
and intelectin in a population of cells \cite{key-11}. In wild-type
and $Nkx\,3.1^{+/-}$ prostate, probasin is expressed uniformly whereas
in the case of the $Nkx\,3.1^{-/-}$ prostate, a heterogeneous population
of probasin-expressing and nonexpressing cells is obtained. In the
case of intelectin expression, a considerable heterogeneity is observed
even in the wild-type prostate. This contrasts with the relatively
uniform expression of the $Nkx\,3.1$ gene in both the wild-type and
$Nkx\,3.1^{+/-}$ prostates. 

\begin{figure}
\begin{center}\includegraphics[%
  width=2.5in]{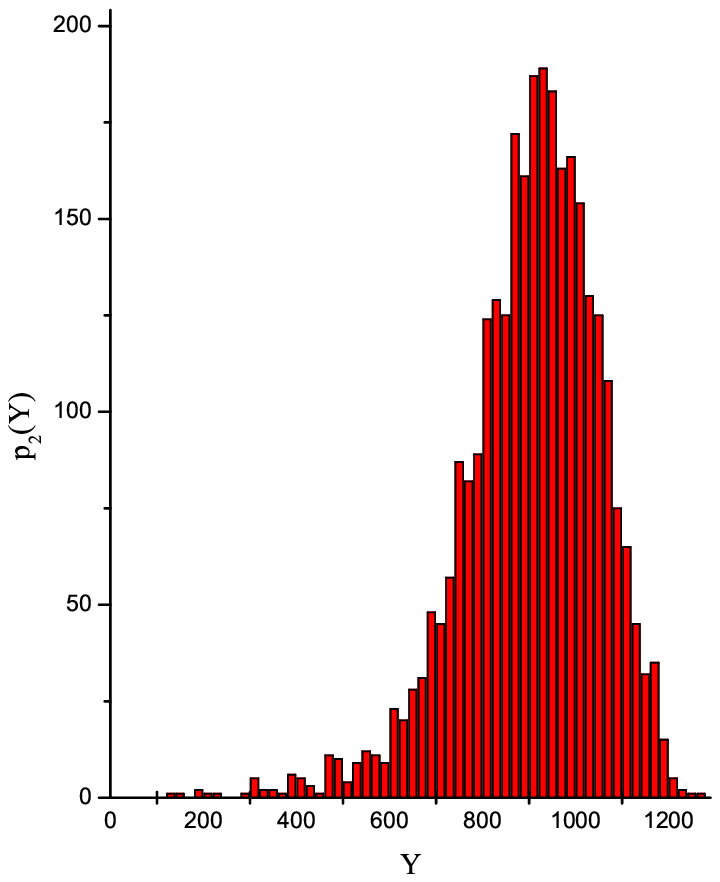} \includegraphics[%
  width=2.5in]{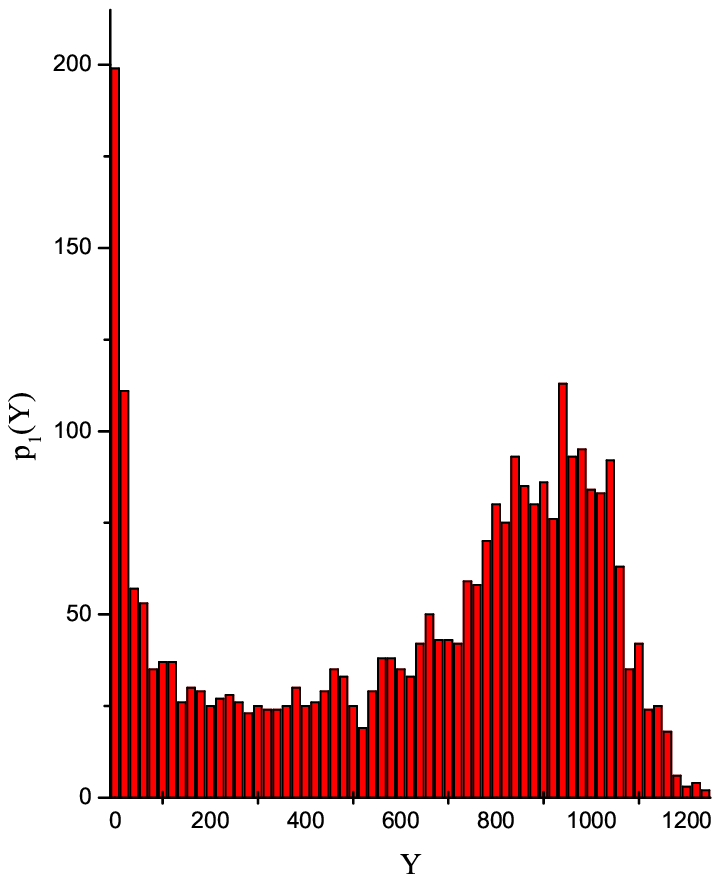}\end{center}

\begin{center}(a)$\qquad\qquad\qquad\qquad\qquad\qquad\qquad$(b)\end{center}

Figure 8. Steady state distribution of protein $Y$ levels, obtained
through numerical simulation based on the Gillespie algorithm, when
the copy number of Gene A is two (a) and one (b) respectively. The
corresponding distributions of protein $Y$ levels are $p_{1}(Y)$
and $p_{2}(Y)$. 
\end{figure}

We now provide an explanation for the dosage response of the probasin
and intelectin genes. The TF ($Nkx\,3.1$ protein)-regulated promoter
activity can be either graded or binary. In the first case, the activity
increases in a graded fashion in response to increasing levels of
TFs. Reductions in the target gene protein levels due to the loss
of one copy of the regulating gene are expected to be uniform in all
cells. In the binary mode, a gene can be in a transcriptionally active
or inactive state and the TFs regulate the probability of the gene
being in either state. Reduced $Nkx\,3.1$ gene dosage leads to a
decrease in the fraction of transcriptionally active cells. At the
population level, a bimodal distribution of the target gene levels
is obtained. In a fraction of cells, the target gene activity is null,
i.e., the signalling pathways leading to the arrest of cell proliferation
or apoptosis remain inactivated. Gene copy reduction combined with
stochastic effects are responsible for the heterogeneous gene activity
in an ensemble of cells.

\begin{figure}
\includegraphics[%
  width=2.5in]{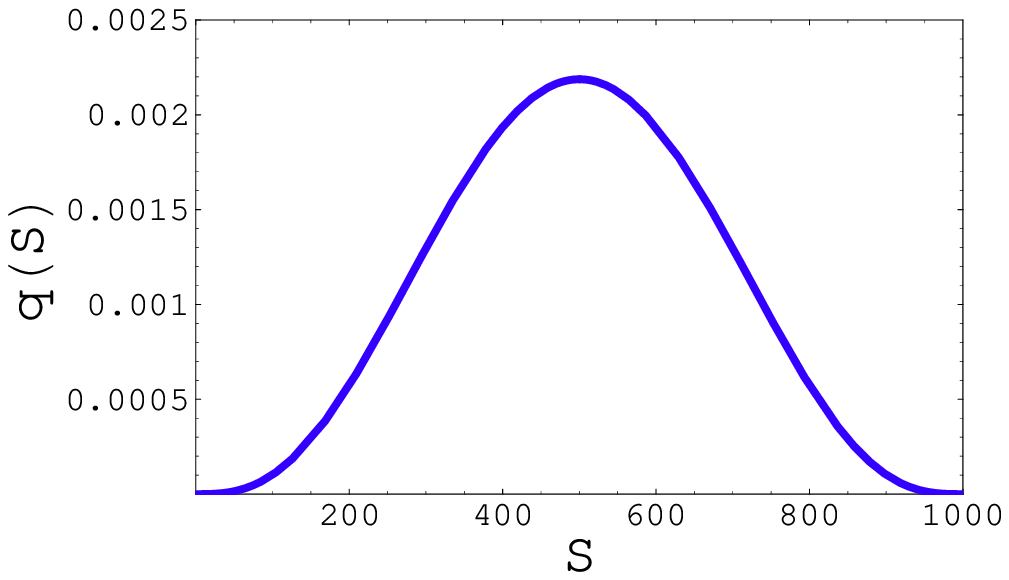} \includegraphics[%
  width=2.5in]{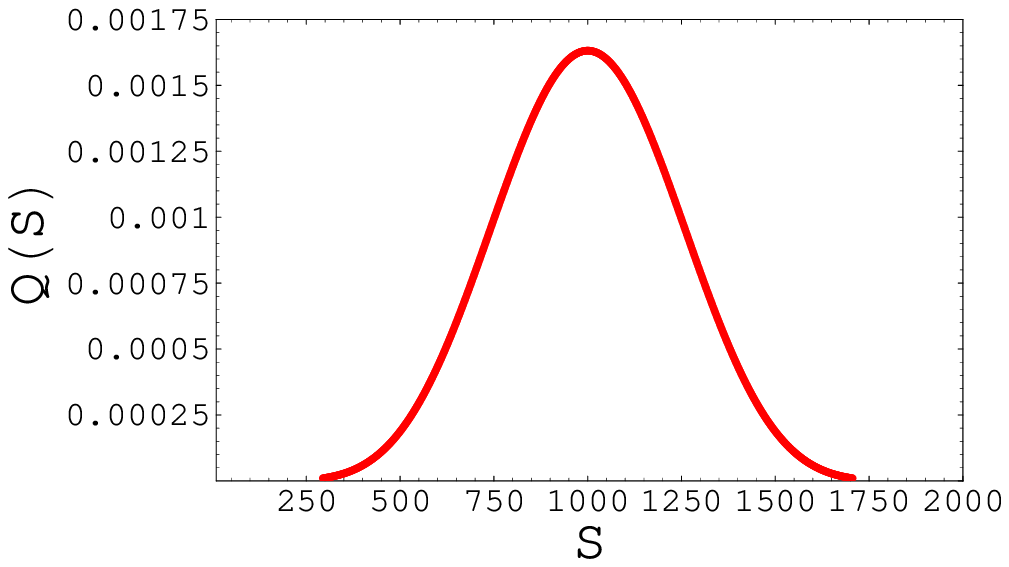}

\begin{center}(a)$\qquad\qquad\qquad\qquad\qquad\qquad\qquad$(b)\end{center}

Figure 9. Steady state distribution of TF levels in the one-gene (a)
and two-gene cases (b).
\end{figure}

\begin{figure}
\includegraphics[%
  width=2.5in]{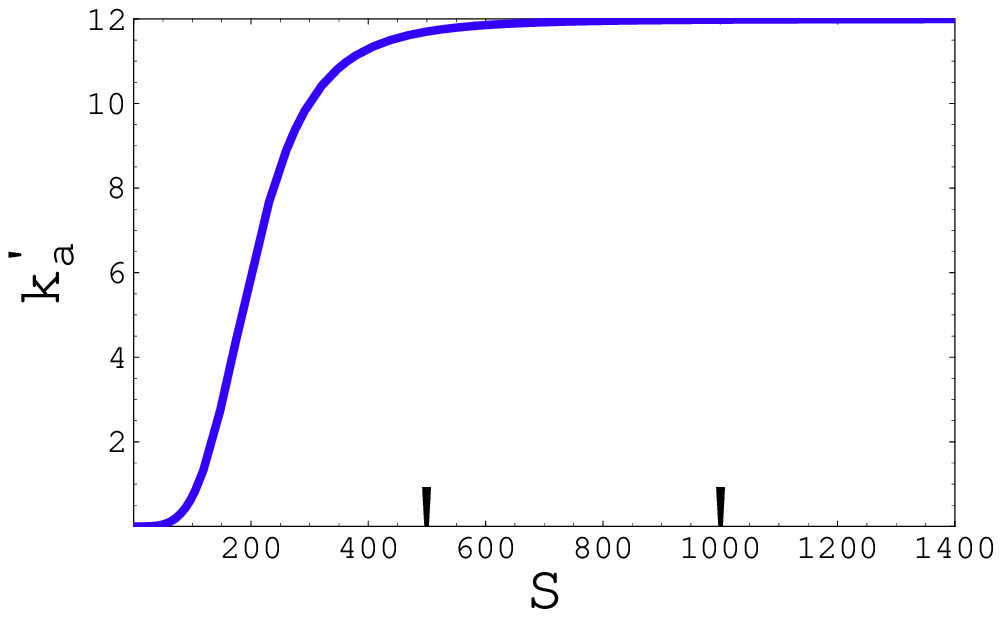} \includegraphics[%
  width=2.5in]{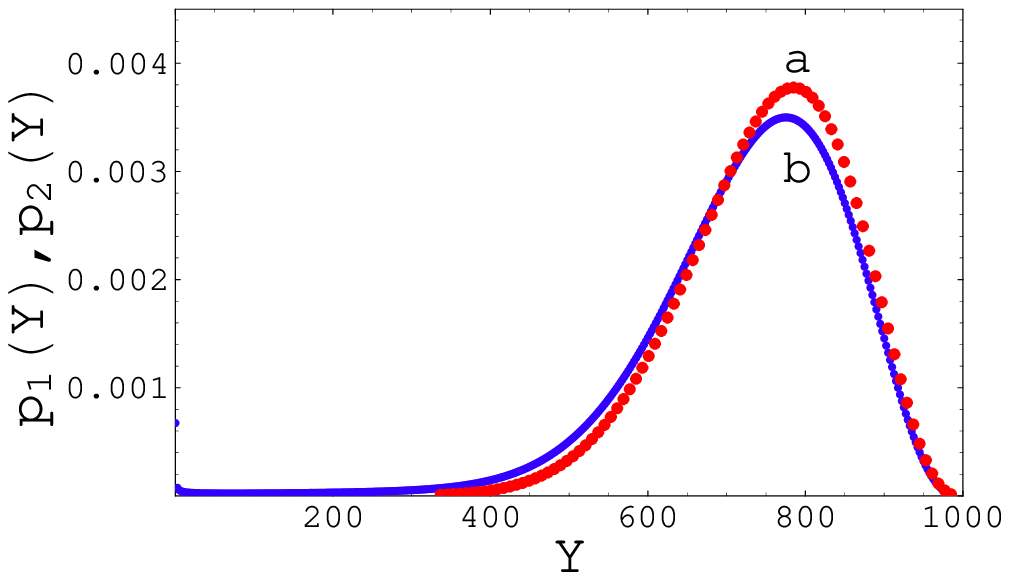}

\begin{center}(a)$\qquad\qquad\qquad\qquad\qquad\qquad\qquad$(b)\end{center}

Figure 10. A possible scenario for probasin gene expression. (a) The
dose-response curve, (b) the distributions $p_{2}(Y)$ ({}``a'')
and $p_{1}(Y)$({}``b'') of target gene protein levels when the
copy number of Gene A is two and one respectively.
\end{figure}

\begin{figure}
\includegraphics[%
  width=2.5in]{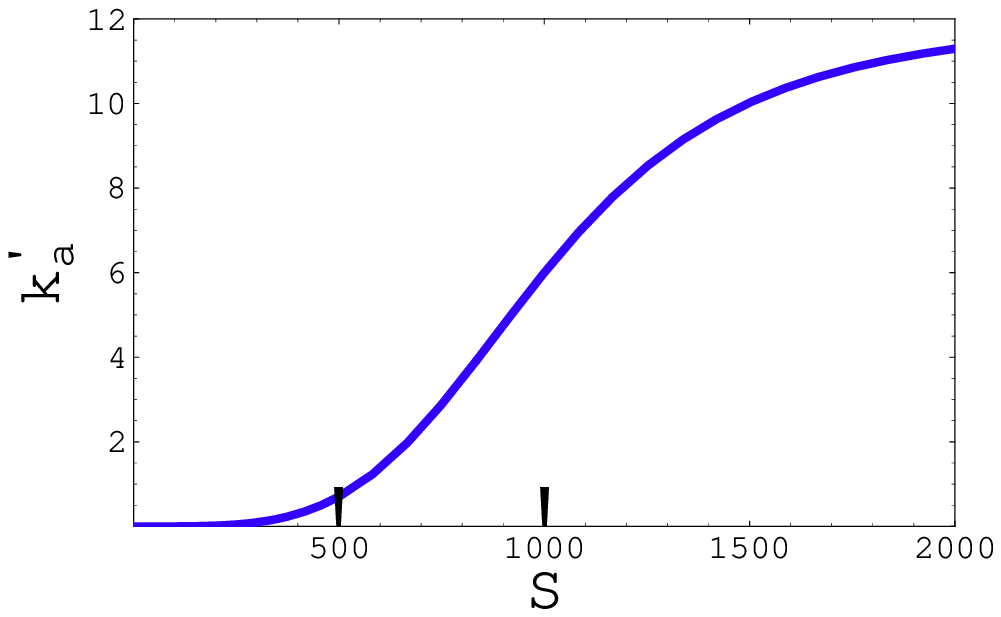} \includegraphics[%
  width=2.5in]{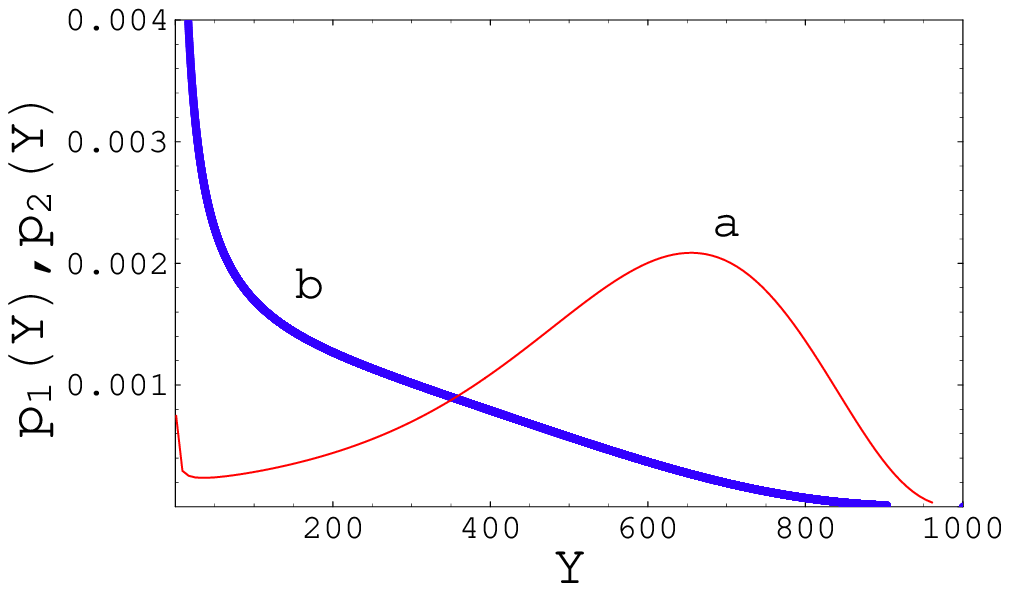}

\begin{center}(a)$\qquad\qquad\qquad\qquad\qquad\qquad\qquad$(b)\end{center}

Figure 11. A possible scenario for intelectin gene expression. (a)
The dose-response curve, (b) the distributions $p_{2}(Y)$ ({}``a'')
and $p_{1}(Y)$({}``b'') of target gene protein levels when the
copy number of Gene A is two and one respectively.
\end{figure}

We now illustrate the experimentally observed responses of the probasin
and intelectin genes to the reduced copy number of the $Nkx\,3.1$
gene in the analytical framework of our stochastic model. Gene A in
the model now represents the $Nkx\,3.1$ gene and Gene B the probasin
or intelectin gene. The particular values of the different rate constants
and the parameters are not mentioned as the aim is to simply demonstrate
how the differences between the responses of the probasin and intelectin
genes arise. Figure 9 shows the distributions $q(S)$ and $Q(S)$
of the TF protein levels when the copy number of Gene A is one and
two respectively. Figure 10(a) shows the dose-response curve $k_{a}^{''}$
versus $S$ of the target gene. One finds that the effective activation
rate constant $k_{a}^{''}$ (Eq. (3)) does not change appreciably
when the copy number of the regulating gene is reduced from two to
one ($S=S_{mean}$ is reduced from 1000 to 500, these two points are
marked by vertical lines on the $x-axis$ of figures 10(a) and 11(a)).
This is reflected in the distributions $p_{1}(Y)$ (curve {}``b'')
and $p_{2}(Y)$ (curve {}``a'') of the target gene protein levels
(Figure 10(b)). Figure 10 represents the case of probasin for which
experimental observations show that the probasin protein level is
relatively insensitive to the loss of one copy of the $Nkx\,3.1$
gene and the probasin gene is expressed uniformly (unimodal distribution
of the protein levels) in both the wild-type and $Nkx\,3.1^{+/-}$
prostate. Figure 11 depicts the possible scenario for the intelectin
gene. In this case, the magnitude of $k_{a}^{''}$ is significantly
reduced (figure 11(a)) when the copy number of Gene A is reduced from
two to one. In the two-gene case, the effective rate constant $k_{a}^{''}$
falls in the most sensitive region of the dose-response curve. The
effect of the TF fluctuations is the most prominent in this case giving
rise to increased heterogeneity in the target gene expression (the
bimodal distribution $p_{2}(Y)$, curve {}``a'', in figure 11(b)).
A bimodal distribution of intelectin-expressing and nonexpressing
cells has been experimentally observed \cite{key-11} in the wild-type
prostate (gene copy number of the $Nkx\,3.1$ gene is two). In the
one gene case, the intelectin expression is found to be essentially
lost. The distribution $p_{1}(Y)$ of the intelectin protein levels
in figure 11(b) (curve {}``b'' ) shows a prominent peak at zero
protein level, in agreement with experimental findings.

\section{Summary and Discussion}

In this paper, we have considered a stochastic model of TF-regulated
gene expression and studied the effect of the TF fluctuations on the
distribution of the target gene protein levels. We have shown that
the TF fluctuations associated with the highest signal sensitive region
of the dose-response curve have the strongest influence on the distribution
of the target gene protein levels, consistent with experimental observations
\cite{key-2,key-3,key-4}. In fact, the TF fluctuations can give rise
to a bimodal distribution in the output protein levels. This is an
experimentally observed effect \cite{key-2,key-3}. The effect of
the TF fluctuations is more prominent for higher values of the Hill
coefficient $n$. The TF fluctuations may be ignored away from the
region of highest signal sensitivity. We have reported the results
for one set of parameter values but the results are of general validity.
The analytical results obtained in the paper are valid when the random
transitions between the active and inactive gene expression states
constitute the dominant source of noise which is often true for eukaryotic
systems. It will be of interest to extend the results of the model
to more general situations. In our analytical formalism, the PDF associated
with the distribution of the TF levels in the steady state is used
to determine the steady state distribution of the target gene protein
levels (see equation (7)). This is similar in spirit to the Static
Noise Approximation studied recently by Scott et al. \cite{key-36}.
The method does not take into account the dynamical aspect of the
TF noise like the frequency of fluctuations in the TF amounts. The
numerical simulation based on the Gillespie algorithm (figure 8) is
more exact in nature and incorporates the dynamical effects.

We have further studied the consequences of reducing the copy number
of Gene A, synthesizing TFs, from two to one. The TF fluctuations
are greater in magnitude when the copy number of gene A is one. As
before, the fluctuations have the strongest effect when the mean TF
protein level, $S_{mean}$ ($n_{G}=1$), is close to $K$, the threshold
parameter for activation of the target gene expression. This is the
region where the dose-response or equivalently the $k_{a}^{''}$ versus
$S$ curve has the highest slope, i.e., maximal sensitivity. As illustrated
in figure 6, the TF fluctuations can give rise to a bimodal distribution
in the output protein levels. The appearance of a bimodal distribution,
when the copy number of the gene synthesizing the TFs is reduced from
two to one, is a prediction of our model and should be verified experimentally.
The result, obtained through analytical calculations, is supported
by the results of numerical simulation based on the Gillespie algorithm.
The variance in protein levels may be underestimated by a significant
factor if transcription and translation are combined into a single
step \cite{key-34,key-22}. In the light of this possibility, transcription
and translation are treated as separate stochastic processes in the
numerical simulation. The use of Hill functions introduces errors
in a stochastic analysis \cite{key-33}. In the simulation, the formation
of a TF complex occurs through stochastic processes. The agreement
between the analytical and numerical results confers validity on the
reported result. 

The issue of whether a protein level exceeds a threshold value is
of crucial importance as the functional activity of proteins, in general,
depends on this. Our study shows that results, significantly altered
from those in the deterministic case, may be obtained when stochasticity
is taken into account. As seen in figure 7, the TF fluctuations can
increase as well as reduce the steady state probability $p(Y>Y_{thr})$,
that the $Y$ protein level exceeds a threshold value $Y_{thr}$,
from the values obtained when the TF fluctuations are ignored. Our
stochastic model enables us to obtain analytical expressions for the
probability distributions. With the knowledge of the distributions,
the calculation of the probability that the target gene protein level
exceeds a threshold value is straightforward. Such calculations provide
the basis for the study of problems related to TF haploinsufficiency.
There are already suggestions in the biomedical literature \cite{key-5,key-10,key-11,key-12,key-13,key-16,key-29}
that stochasticity is a key contributing factor in the occurrence
of haploinsufficiency. Our study lends support to this notion and
shows that the gene copy number is an important criterion in determining
the relationship between stochasticity and desired protein activity.
We have specifically considered the case of $Nkx\,3.1$-regulated
gene expression in the prostate. The loss of one copy of the $Nkx\,3.1$
gene can trigger the initiation and growth of tumour leading to cancer.
Our stochastic model of TF-regulated gene expression provides an explanation,
at a qualitative level, for the experimental observations on the target
gene responses to gene (TF) dosage reduction. A recent experiment
\cite{key-25} shows that the loss of one allele of the $p53$ tumour
suppressor gene results in a four-fold reduction of $p53$ mRNA and
protein, thus providing the basis for haploinsufficiency. As in the
case of the $Nkx\,3.1$-regulated probasin and intelectin genes, the
$p53$-dependent transcriptional response after genetoxic stress can
be either homogeneous or heterogeneous \cite{key-26}. A heterogeneous
response (only some cells of a population respond, the fraction of
such cells increases as the TF amount increases) can be explained
only if stochastic gene expression is taken into account. A reduction
in the gene copy number accenuates the stochastic effects. Investigation
of the stochastic origins of haploinsufficiency is expected to yield
useful information and insight on a host of human diseases. 

\begin{center}\textbf{Acknowledgement}\end{center}

R.K is supported by the Council of Scientific and Industrial Research,
India under Sanction No. 9/15 (239)/2002-EMR-1.

\end{document}